\documentclass[%
 aip,
 amsmath,amssymb,
 reprint,
]{revtex4-2}

\usepackage{graphicx}
\usepackage{dcolumn}
\usepackage{bm}

\usepackage[utf8]{inputenc}
\usepackage[T1]{fontenc}
\usepackage{mathptmx}
\usepackage{etoolbox}
\usepackage{placeins}
\usepackage{nameref}
\usepackage{enumitem}
\makeatletter
\def\@email#1#2{%
 \endgroup
 \patchcmd{\titleblock@produce}
  {\frontmatter@RRAPformat}
  {\frontmatter@RRAPformat{\produce@RRAP{*#1\href{mailto:#2}{#2}}}\frontmatter@RRAPformat}
  {}{}
}\makeatother%

\begin{document}

\preprint{AIP/123-QED}

\title{Automated Tabletop Exfoliation and Identification of Monolayer Graphene Flakes}
\author{E.D.S. Courtney}
\affiliation{Stanford Institute for Materials and Energy Sciences, SLAC National Accelerator Laboratory, Menlo Park, CA 94025}
\affiliation{ 
Department of Physics, Stanford University, Stanford, CA 94305
}%
\email{goldhaber-gordon [at] stanford [dot] edu}
\email{elijahc [at] stanford [dot] edu}
\author{M. Pendharkar}%
\affiliation{Stanford Institute for Materials and Energy Sciences, SLAC National Accelerator Laboratory, Menlo Park, CA 94025}
\affiliation{Department of Materials Science and Engineering, Stanford University, Stanford, CA 94305}
\author{N.J. Bittner}
\affiliation{Independent Researcher}
\author{A.L. Sharpe}
\affiliation{Materials Physics Department, Sandia National Laboratories, Livermore, CA 94550}
\author{D. Goldhaber-Gordon}
\affiliation{Stanford Institute for Materials and Energy Sciences, SLAC National Accelerator Laboratory, Menlo Park, CA 94025}
\affiliation{ 
Department of Physics, Stanford University, Stanford, CA 94305
}%

\date{\today}

\begin{abstract}
Over the past two decades, graphene has been intensively studied because of its remarkable mechanical, optical, and electronic properties. Initial studies were enabled by manual ``Scotch Tape'' exfoliation; nearly two decades later, this method is still widely used to obtain chemically-pristine flakes of graphene and other 2D van der Waals materials. Unfortunately, the yield of large, pristine flakes with uniform thickness is inconsistent. Thus, significant time and effort are required to exfoliate and locate flakes suitable for fabricating multilayer van der Waals heterostructures.
Here, we describe a relatively affordable tabletop device (the ``eXfoliator'') that can reproducibly control key parameters and largely automate the exfoliation process. In a typical exfoliation run, the eXfoliator produces 3 or more large ($\ge400\ \mu$m$^2$) high-quality graphene monolayer flakes, allowing new users to produce such flakes at a rate comparable to manual exfoliation by an experienced user. We use an automated mapping system and a computer vision algorithm to locate candidate flakes. Our results provide a starting point for future research efforts to more precisely identify which parameters matter for the success of exfoliation, and to optimize them.
\end{abstract}

\maketitle

\section{Introduction}

Monolayer graphene, a single layer of carbon atoms arranged in a honeycomb lattice with an atomic spacing of 1.42\AA, has remarkable mechanical, optical, and electronic properties. Graphene has recently emerged as an exciting basis for studying strongly-correlated electron physics. Common stacking orders of multi-layer graphene, such as Bernal bilayer and rhombohedral trilayer and pentalayer can display superconductivity \citep{Young2021,Young2022,Wang2019} and orbital magnetism\citep{Wang2022,Ju2023}. In addition to these configurations exfoliated from natural graphite, monolayer graphene can be artificially stacked with interlayer twists to form moir\'e materials with a large structural design space, including configurations with flat electronic bands and non-trivial band topology. Superconductivity and orbital magnetism\citep{Cao2018,Sharpe2019}, and fractional Chern insulators\citep{Yacoby2021,Han2023}, have been discovered in various graphene-based moir\'e materials. Motivated by the rich set of target structures based on this single material, we focus our discussion and experimental demonstration on graphene, but we anticipate our methods and workflow will apply more broadly to the burgeoning family of exfoliatable van der Waals layered materials.

In the Scotch tape exfoliation method introduced by Geim and Novoselov in 2004\citep{Nov2004}, natural or man-made graphite crystals –- which may consist of $10^4$--$10^6$ graphene layers arranged in large crystallographic domains –- are repeatedly adhered to adhesive tape and cleaved apart to populate large areas of the tape with increasingly thin graphite crystals.
The graphite-populated surface of the tape is then applied to a Si substrate coated with a thin SiO${_2}$ layer, before the tape is peeled away.
Graphene flakes with a distribution of thicknesses ranging down to just one layer cleave from the graphite and are left on the substrate.

Other techniques for producing large-area monolayers from bulk graphite, including electrostatic exfoliation, gold-tape exfoliation, intercalation, and sonication continue to be developed, and graphene can also be grown over large areas by chemical vapor deposition (CVD)\citep{Sidorov2007,Liu2021,Luo2022,Amontree2024}. Each of these techniques has its own advantages, as discussed below, but Scotch tape exfoliation remains pre-eminent when seeking low disorder graphene-based structures in a research -- as opposed to industrial production -- environment.

Electrostatic exfoliation uses high potentials between two metallic electrodes to directly cleave graphite into mono- and few-layer graphene flakes, but often results in smaller flakes with higher numbers of defects compared to flakes obtained by tape exfoliation, especially in monolayers\citep{Sidorov2007}.

Metal-assisted or metal-tape exfoliation can produce large monolayers of transition metal dichalcogenides materials or graphene, starting from a bulk crystal or an epitaxial thin film\citep{Kim2018,Huang2018,Liu2021}.
These techniques depend on strongly adhering the target monolayer to a metal film. The monolayer is subsequently transferred to a new substrate and the metal film chemically etched away. Questions about whether metal and etchant can be fully removed without damaging the monolayer film have so far limited adoption of these approaches in research aimed at correlated electron states. 
Sonication and intercalation to extract monolayer graphene produce large numbers of flakes but leave adsorbates on the graphene and often produce flakes smaller than 1 micron\citep{Liu2021,Cies2014}.

Large-area growth techniques have produced promisingly large single-crystal graphene monolayers on a variety of substrates\citep{Lee2014,Wang2014,Luo2022,Amontree2024}. Transfer of these monolayers to silicon has typically involved solvents and sometimes etchants. Dry transfer, most suitable for seeking exotic electronic states, has been rare for these grown films; despite notable successes in oxidizing a copper growth substrate then picking up graphene using exfoliated hBN\citep{Banszerus2015}, this method has not been widely adopted.

Some of these techniques are popular in particular contexts, but as noted above tape exfoliation remains dominant in experiments aimed at studying correlated electronic states. We set out to design a tabletop machine to reduce the time researchers must spend on tape exfoliation and to make the process’s yield more consistent. Our metrics for success include the production rate, number density, size, and disorder of flakes deposited.

We hypothesized that if the key parameters of the exfoliation process could be identified and controlled, optimal values could be determined through experiments using a purpose-built tool, and then reproduced in successive exfoliations with that tool. When surveyed, researchers in our lab reported that manually exfoliating graphene on Si/SiO$_2$ chips, including the initial tape population, takes between 1 and 6 minutes per square centimeter of exfoliated area. This time depends on an individual researcher's chosen recipe, which can vary in the area prepared, technique for populating the tape, technique for pressing tape against substrate, anneal temperature and duration, and peel speed and angle. Given the anecdotally reported variation in both exfoliation technique and yield, we hypothesized that translating exfoliation to a semi-automated standardized process run on a machine could enable a researcher to produce more large single-crystal freestanding monolayers than a series of manual exfoliations in a comparable amount of time. 
The search for good process parameters need only be carried out once for each material, so any operator might be able to achieve consistent results after a shorter learning period than that required for manual exfoliation. A machine could also reach process parameters outside the range that a human researcher could readily and consistently implement manually (e.g., peeling at 0.1 mm/s or maintaining a chip at 80$^\circ$C during peeling), possibly enabling a machine to achieve better results than manual exfoliation in the hands of even an experienced practitioner. Finally, such a machine would free the operator to pursue other work while running an otherwise tedious or time-consuming exfoliation recipe.

In fact, based on similar considerations, others have already implemented automated exfoliation systems.
A prominent example of automated exfoliation is Brookhaven's Quantum Material Press (QPress)\citep{QPRESS}, a sophisticated cluster of almost fully automated tools capable of exfoliating, cataloguing, stacking, and storing 2D materials in an inert atmosphere.
QPress has successfully automated most of the tape exfoliation process, is accessible through user proposals, and may eventually provide flakes to many collaborators. In the opposite limit, a simple automated exfoliation tool can consist of a single linear motor integrated with a hot plate\citep{Wang2022email} to control peel speed and anneal temperature.

The tool presented in this paper (the ``eXfoliator'') was designed to fill a middle ground, as a user-friendly machine capable of controlling a range of likely-relevant parameters during exfoliation, balancing utility with simplicity of design and construction.

The parameters of Scotch tape exfoliation targeted by the design are peel angle, peel rate, application pressure, temperature of the substrate at all stages, and dwell time under pressure or at temperature. Other relevant parameters, such as preparation of the SiO$_2$ surface by oxygen plasma cleaning, are known to have a significant effect but are not explicitly controlled by the eXfoliator\citep{Huang2015}.

Estimates of graphite's interlayer van der Waals binding energy vary, but most calculated values for AB, AAA, and ABC-stacked graphite range from 40-60 meV/atom\cite{Chen2013}. For comparison, the thermal energy of graphite at 300 (400) K is approximately 1300 (2400) J/mol, corresponding to 13.5 (25) meV/atom\citep{Pop2012}. Because the thermal energy at accessible temperatures -- as limited by the melting point of the tape's backing -- does not approach the binding energy threshold, we assume tuning the annealing or peeling temperature primarily affects the viscoelastic behavior of the polymer tape adhesive.

From the time-temperature superposition principle of polymer flow\citep{Li2000}, we believe adjusting the annealing temperature, application pressure, or dwell time allows the viscoelastic polymer tape adhesive to flow more freely over the graphite flakes and allow better contact before being cooled and peeled. Likewise, we assume adjusting peel speed tunes the viscoelastic behavior of the tape adhesive instead of significantly affecting the fundamental interlayer interactions of the graphene layers. Adjusting the peel angle is believed to change both the radius of curvature of the tape and the ratio of shear to normal forces at the interlayer boundary.

Graphene is typically manually exfoliated onto substrates with an area of one to a few cm$^2$ cut from a Si/SiO$_2$ wafer.
The eXfoliator exfoliates over a 50x50 mm square (25 cm$^2$ area, 5-20x that in a typical manual exfoliation) on a 76 mm diameter Si/SiO$_2$ wafer.
To take advantage of this increased scale we purchased a microscope with a motorized stage capable of imaging contiguous areas as large as 8 cm on a side, and we developed a computer-vision algorithm to reliably identify images containing monolayer graphene flakes. The algorithm empirically relates local substrate color to monolayer appearance, and identifies contiguous clusters of pixels, allowing efficient evaluation of exfoliation output and reducing both labor and time required to locate the largest monolayers on the substrate. The imaging and flake identification will be described elsewhere, but similar techniques have been previously demonstrated\citep{Masubuchi2018,Uslu2024}.

\section{Materials and Methods}
The Scotch tape exfoliation process described here entails creating graphite-covered tape, applying the populated tape to an oxidized silicon substrate, annealing that substrate, and subsequently peeling the tape from the silicon. The silicon substrate must then be searched to locate resulting monolayers, which can be picked up and used to create graphene-based devices. 

\subsection{The eXfoliator}
\begin{figure}
    \centering
    \includegraphics[width=0.45\textwidth]{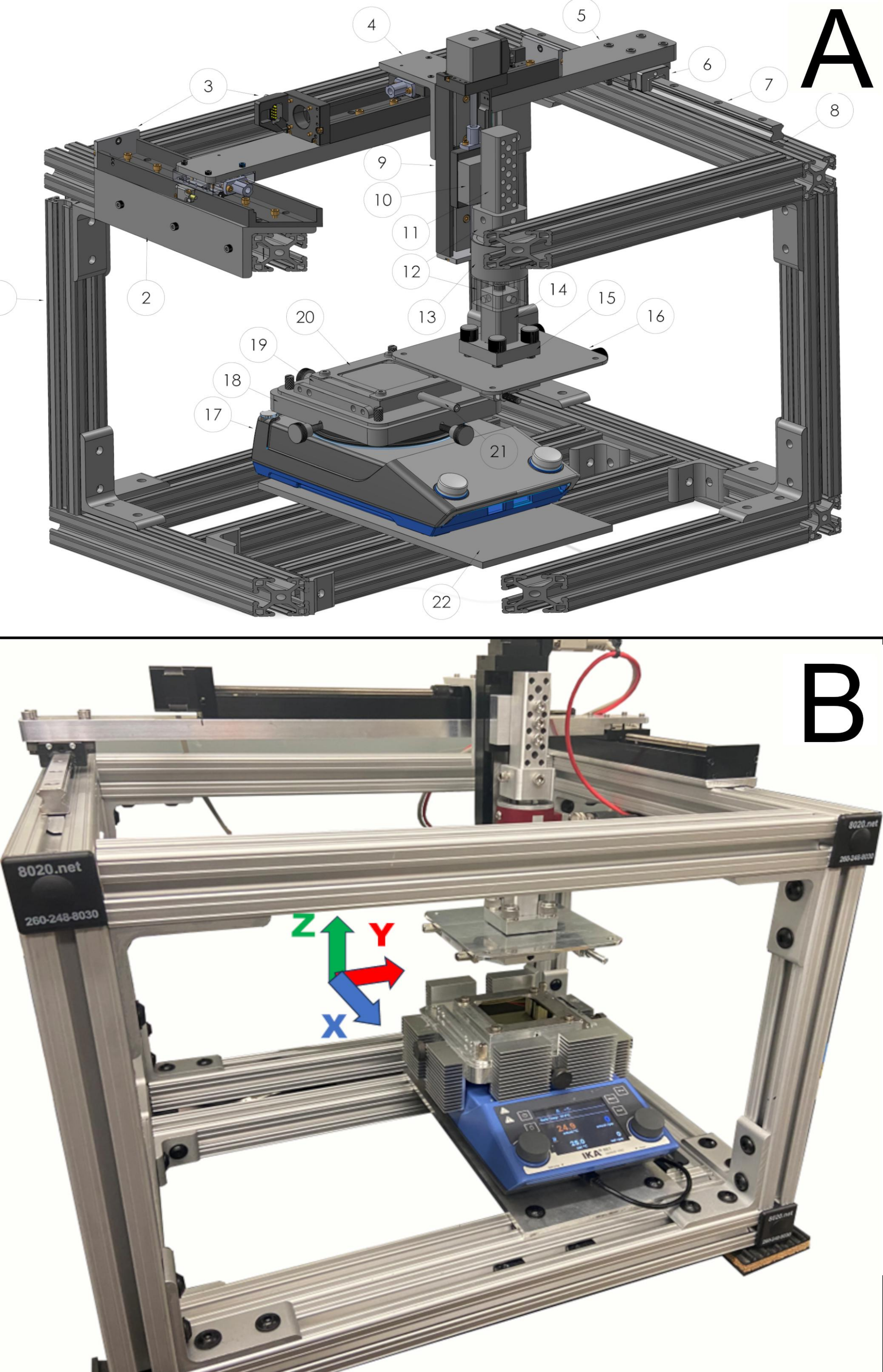}
    \caption{A. A schematic of the eXfoliator with key parts labeled; see Appendix A for the parts list. CAD files are provided through the Stanford Digital Repository\citep{exfolfiles}. B. The constructed eXfoliator, with populated tape prepared to be pressed to a silicon wafer. The peeling motion takes place in the Y-Z plane. Aluminum fin comb heatsinks were added to the hot plate to speed cooling with the aid of a small external fan (not shown).}
    \label{fig:realimg}
\end{figure}

A schematic of the eXfoliator is shown in Fig. \ref{fig:realimg}. The figure caption points the reader to descriptions and CAD files for all parts. Three perpendicularly-oriented linear stepper motors (Fig. \ref{fig:realimg}B, Parts 3,9) mounted on a metal frame (Parts 1-2,4-8) to move an extended arm (Parts 10-16) above a remotely controlled hot plate (Part 17) with a custom machined aluminum working surface (Parts 18-21). The machine itself occupies a nearly cubic volume of 46x54x51 cm. The motors, force gauge, and hot plate are connected to an external computer running a custom GUI that enables either manual control or scripting with recipes that can be standardized.

We selected the hot plate based on its computer control interface and its built-in weighing scale to measure the force with which the tape was being pressed onto the substrate; however, the weighing function proved unreliable, so we incorporated a separate force gauge. Adding a plate adapter and wafer pedestal (Parts 18-19) substantially increased the time needed to reach temperature set points. For future implementations of the eXfoliator design, we suggest considering other hot plates with increased heating power to more quickly reach temperature set points; a minor redesign of Part 18 would be required to accommodate a different working surface.

As shown in Figure \ref{fig:realimg}B, we added fins to speed cooling, aiming to approximate a common method of manual exfoliation where the tape-silicon sandwich is removed from the hot plate and allowed to cool within seconds prior to peeling. However, even the modified version took nearly 25 minutes to cool from 105$^\circ$C to below 30$^\circ$C; incorporating active cooling may prove beneficial.

\subsection{Graphene Exfoliation}
In Scotch tape exfoliation, natural or synthetic graphite crystals are repeatedly adhered to adhesive tape and cleaved apart to populate a large area of the tape with successively thinner graphite crystals; as this process roughly doubles the graphite-populated area of tape with each cleave, a 25 cm$^2$ area may be fully populated in as few as 8 cleaves from a 9 mm$^2$ starting crystal. 
The eXfoliator was designed to be able to populate two 25 cm$^2$ areas of tape simultaneously, but the process proved inefficient compared to manually populating the tape.
Therefore, for all results reported here, we manually populated an initial `mother tape', which we then used to populate a 25 cm$^2$ `daughter tape' to be used in the eXfoliator, a process analogous to that commonly used in manual exfoliation. The materials used by the eXfoliator in this paper are also commonly used in manual exfoliation: 1" Scotch magic tape, ``flaggy'' graphite flakes purchased from NGS Trading \& Consulting, and 76mm silicon wafers topped with 300nm wet silicon oxide purchased from NovaWafers.
However, the eXfoliator's design does not limit it to these materials. We envision that the eXfoliator can be easily used to exfoliate most vdW materials, including hBN and transition metal dichalcogenides, onto a wide variety of substrates using a wide variety of adhesive surfaces to hold the initial material.

During operation, a pristine 76 mm Si/SiO$_2$ wafer is clamped in place on the wafer holder (Part 19) by Part 20. The tape plate (Part 16) is placed on a bench top with the tape-mounting surface face-up. Two side-by-side 1’’ strips of tape are mounted, sticky side up, to the tape plate. These serve as the daughter tape. The daughter tape is populated by gently pressing the mother tape to one half, before manually peeling it at a steep angle (typically, >120$^\circ$); the process is then repeated with the same mother tape on the other half. The tape plate is then mounted to the moving arm using Part 15, with four intervening parallel wave springs to increase uniformity of contact as pressure is applied. The tape plate is positioned over the target substrate using the GUI, and the tape can then be applied to the substrate at a desired pressure, $P$, annealed at a desired temperature, $T$, and peeled at a desired linear speed $V$, at an angle $\theta$ in the Y-Z plane, defined as the angle formed by the tape with respect to the Y-axis as the tape is peeled from the substrate, assuming the tape remains unstretched. We have used a range of tape application temperatures and peel speeds, but targeted an initial application pressure of 30 kPa, releasing one side of the tape and removing the applied pressure as the wafer cooled from the set temperature to within 5$^\circ$C of the desired peeling temperature of 25$^\circ$C. The linear motion of the plate during peeling resulted in the peel angle varying continuously from 60$^{\circ}$ to 80$^{\circ}$ over each exfoliation, as calculated from the geometry of the system.

\begin{table}[h]
    \centering
    \caption{Exfoliation parameters considered in this study, with the extreme values accessible by eXfoliator, typical values used in this study, and representative values for manual exfoliation.}
    \label{tab:paramtable}
    \begin{tabular}{|c|c|c|c|c|c|}
    \hline
        Parameter & Min & Max & Typical & Manual \\
        \hline
         $P$ & 400 Pa & 40 kPa & 30 kPa & 1-10 kPa\\
         \hline
         $T$ & 25$^{\circ}$C & 200$^{\circ}$C & 25-100$^{\circ}$C & 90$^{\circ}$C\\
         \hline
         $V$ & 1 $\mu$m/s & 1.2 cm/s & 1 $\mu$m/s-1 cm/s&1 mm/s\\
         \hline
         $\theta$&60$^{\circ}$&170$^{\circ}$&60-80$^{\circ}$&30-170$^{\circ}$\\
         \hline
    \end{tabular}
\end{table}
After the tape is peeled, the wafer with exfoliated graphene is transferred to the motion stage of a Leica DM6 M microscope.
The area on the wafer where exfoliation is performed is then imaged in an array of photos through a 10x objective. Exposure, gain, white balance, gamma, color saturation, and illumination intensity are kept constant. A flat-field correction equalizes brightness across each image. A focus map for the scan area is made using between 10 and 50 manually determined XYZ values as focal points across the wafer's surface. Scanning each wafer generates approximately 3200 24-bit 20MP images to be analyzed by a flake-finding algorithm we designed for this purpose. Though we designed this algorithm independently, we later encountered a similar but more sophisticated method in the recent literature\citep{Uslu2024}.

\subsection{Wafer Imaging and Monolayer Identification}
It is well known that the optical transmission of monolayer graphene is an approximately constant 97.7\% for wavelengths in the visible range\citep{Nair2008}. However, beginning with Geim and Novoselov's work\citep{Nov2004}, most researchers exfoliate on silicon wafers with thin oxide to improve contrast, as the reflection from the underlying silicon roughly doubles apparent contrast. In addition, thin-film interference effects from the oxide layer, typically chosen to be about 90 or 300nm thick, further enhance visibility, enabling monolayer graphene to be recognized in standard optical microscope images without heavy image processing.

Choosing lighting conditions and camera settings to further enhance contrast can help in automating flake detection. All our imaging uses the provided LED illumination of the Leica DM6 M and a 10x magnification objective (Leica Objective \#566503, NA 0.3). In software, we apply a flat-field correction to all images to remove vignetting, and we choose a suitable white-balance which we hold constant through the study. Thus all images are saved in false-color, with enhanced contrast for monolayers against the background appearance of the oxidized silicon.

Though all sample wafers have nominally uniform 300nm thick layers of silicon oxide, we noted that the manufacturer's specified tolerance of $\pm$ 5\% oxide thickness resulted in varying background colors within individual wafers and from wafer to wafer. Thus, we had to determine empirical relations between apparent background and monolayer color across a range of backgrounds. To determine these relations we collected over 200 monolayer-background color pairs from the output of a less accurate algorithm across several samples. We held the relations fixed throughout the study.

The computer vision (CV) algorithm detects the background color in each image, and uses the previously calculated empirical linear relations between background and monolayer colors in the R, G, and B channels under the established lighting settings to predict monolayer appearance. Clusters of pixels within a certain Euclidean distance of the predicted flake color are then identified and labeled, with the corresponding image saved for subsequent review.

\begin{figure}[hbp]
    \centering
    \includegraphics[width=0.45\textwidth]{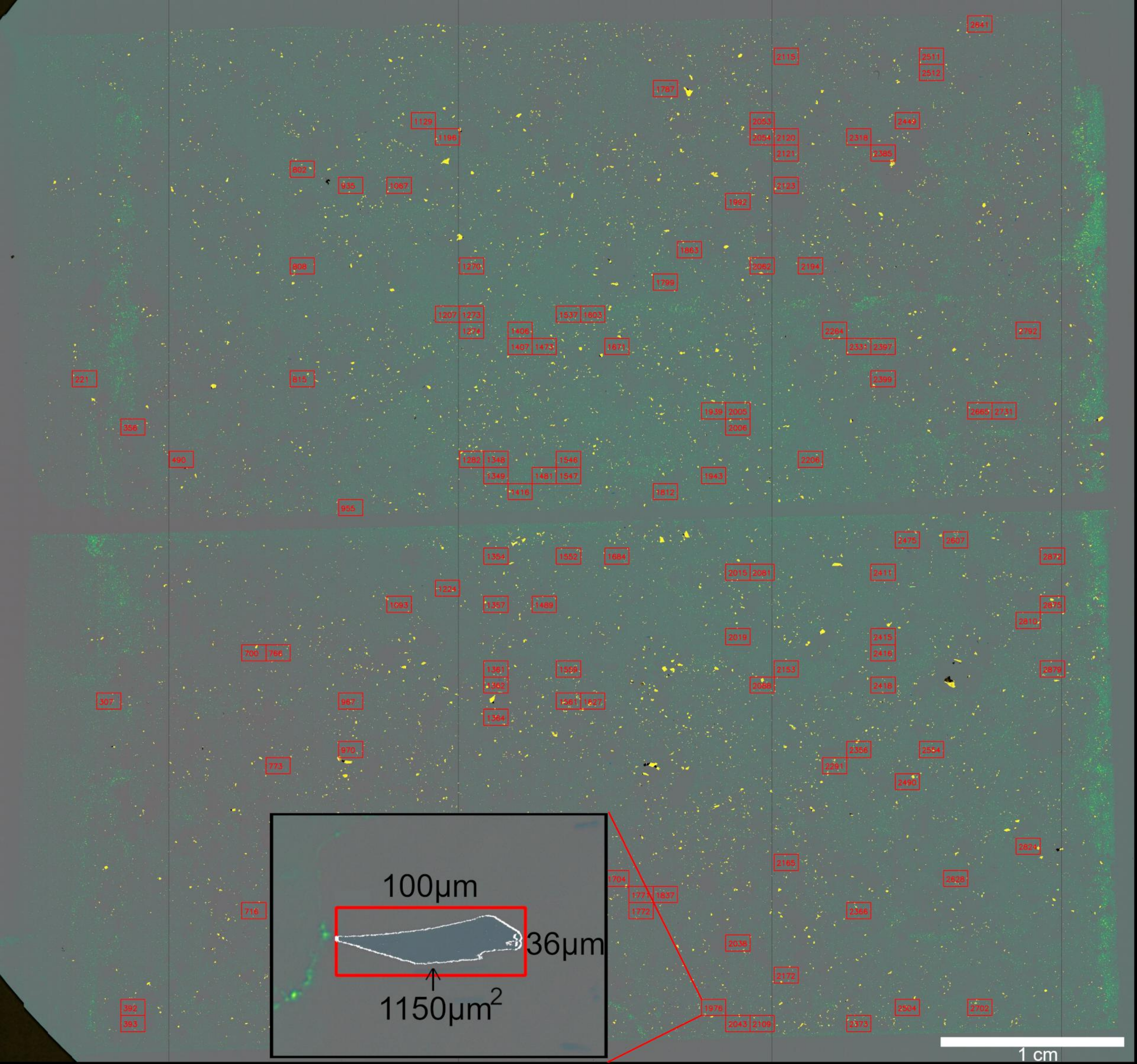}
    \caption{A composite, false-color image of the scanned area on an exfoliated wafer, with identified monolayer locations highlighted in red and labeled according to the image's location in the wafer scan. Scale bar corresponds to 1 cm. This wafer was annealed at 25$^\circ$C and peeled at 100 $\mu$m/s with a contact pressure of 29 kPa and a peel angle swept from 60$^\circ$-80$^\circ$. Inset: a monolayer graphene flake with a calculated area of 1150$\mu$m$^2$.}
    \label{fig:coordmap}
\end{figure}

All images identified as containing monolayers are manually categorized to identify freestanding monolayers and eliminate false positives before being uploaded to a shared repository for use in device fabrication; a wafer-scale stitched image is generated for each sample. Flake locations are labeled to facilitate retrieval, as shown in Figure \ref{fig:coordmap}.
\FloatBarrier
As discussed in Results, the algorithm was less reliable in detecting monolayers with areas below 100 $\mu$m$^2$ - corresponding to a smallest length dimension of 5-10 pixels. However, the most desirable monolayers for assembling TBG stacks are typically above 400 $\mu$m$^2$ in area. We accordingly chose 200 $\mu$m$^2$ as the cutoff value above which we presumed all flakes were found. In addition to this dependence on area, other significant limitations of the algorithm include sensitivity to lighting settings and a failure to detect monolayers located on backgrounds outside the range used to determine the initial empirical relations. Prepared samples were excluded from the study if large discolored areas with no detected monolayer flakes were present while other regions of the sample exhibited a significant distribution of monolayer flakes, as this indicates a failure of the flake-finding algorithm due to background color. However, as the oxide thickness (and resulting background color) is not significantly affected by any step of the exfoliation process (including brief atmospheric anneals up to 105$^\circ$C), this is not expected to bias the reported results.
For a similar, more sophisticated flake-finding algorithm, see Uslu et al., 2024\citep{Uslu2024}.

\section{Results}
In total, we created 33 sample wafers over a range of peel speeds and anneal temperatures, as shown in Table \ref{tab:resultstable1}. In each case we applied the populated tape with an average pressure $P=29\pm 2$ kPa, and peeled with an angle $\theta$ that swept from 60$^\circ$ to 80$^\circ$ across each wafer.


This study was initially aimed to measure the dependence of monolayer yield on peel speed and application temperature, but it has become clear that yield's dependence on uncontrolled variables such as the population of the initial tape, ambient humidity, and substrate contamination is comparable to its dependence on the controlled parameters. A full study of the influence of these parameters likely requires more precise control of at least the initial population step. Though the eXfoliator provides the minimum platform required for such an experiment, a full study is beyond the scope of this work.

In fact, a satisfying outcome is finding a point in the space of controlled parameters that produces large, freestanding monolayers at a rate comparable to manual exfoliation and sufficient to feed the pipeline of device fabrication in the lab, with low levels of polymer residue.

\FloatBarrier
\begin{table}[h]
    \centering
    \caption{Number of sample wafers exfoliated at each peel speed and anneal temperature, with application pressure, dwell time, peel angle, and peel temperature held constant.}
    \begin{tabular}{|c|c|c|c|c|c|}
    \hline
         Count (N)&  1 $\mu$m/s& 10 $\mu$m/s &  100 $\mu$m/s&1 cm/s& Total \\
        \hline
        25$^{\circ}$C & 2 &7 & 8 & 4&21\\
         \hline
         65$^{\circ}$C & 4 & 1 & 2 & 1&8\\
         \hline
         105$^{\circ}$C & 1 & 1 & 1&1&4\\
         \hline
         Total&7&9&11&6&33\\
         \hline
    \end{tabular}
    \label{tab:resultstable1}
\end{table}
\FloatBarrier
After exfoliation, the CV flake searching algorithm was applied and the freestanding monolayers of area greater than 200 $\mu$m$^2$ (``useful'' monolayers) tabulated, with average yields per wafer shown in Table \ref{tab:suppresultstable2} (see Appendix B). 

As the graphite crystals used in this experiment were all formed by natural processes, the size of crystal domains therein might be expected to vary. However, even in the unlikely event that each layer in an initial graphite flake were a single crystal domain, repeatedly cleaving a flake via adhesion to tape during the population step would naturally lead to a distribution of flake dimensions. This can be seen by considering a simple model of an $N$-layer crystal as cleaving into fractions of thickness $\alpha N$ and $(1-\alpha) N$ for some positive constant $\alpha<1$. Using this model, upon repeated cleaving steps - as occur during tape population - the probability $\mathcal{P}_k$ to find a flake of thickness $t_k=N\alpha^{k}(1-\alpha)^{n-k}$ after $n$ cleaves is given by  $\mathcal{P}_k=\frac{1}{2^n}\frac{n!}{k!(n-k)!}$, where values of $t_k<1$ represent monolayers with areas smaller than the starting crystal. This inevitable distribution of flake thickness (and monolayer areas), further compounded by differences in flake thickness and crystal domain size in the starting graphite, may account for some of the variance seen in the observed yields.

To account for this inherent variance in the manual population step, the useful monolayer counts were then normalized by the area of graphite remaining on the populated tape post-exfoliation as a proxy for the initial graphite area. As shown in Table \ref{tab:resultstable3}, slower peel speeds tend to yield a greater number of useful monolayers. 
\begin{table}[h]
    \centering
        \caption{Number of useful monolayers per square mm of graphite on the initial tape at each parameter.}
    \begin{tabular}{|c|c|c|c|c|c|}
    \hline
         UML/A$_{graphite}$ &  1 $\mu$m/s& 10 $\mu$m/s &  100 $\mu$m/s&1 cm/s& Mean \\
         (N/mm$^2$)&&&&&\\
        \hline
        25$^{\circ}$C & 1.7$\pm$0.1 &2.4$\pm$1.6 & 1.4$\pm$0.9 & 0.7$\pm$1.2&1.6$\pm$1.3\\
         \hline
         65$^{\circ}$C & 3.4$\pm$0.8 & 2.9 & 2.6$\pm$1 & 1.8&2.8$\pm$0.9\\
         \hline
         105$^{\circ}$C & 2.9 & 0.9 & 1.6&2.2&1.6$\pm$0.9\\
         \hline
         Mean&2.8$\pm$1&2.2$\pm$1.5&1.5$\pm$1&1.1$\pm$1&1.8$\pm$1.3\\
         \hline
    \end{tabular}
    \label{tab:resultstable3}
\end{table}
\FloatBarrier 
As shown in Tables \ref{tab:resultstable6} and \ref{tab:suppcorrelations}, no clear relationship between peel speed or anneal temperature and individual monolayer area is evident in this experiment. 
\begin{table}[h]
    \centering
        \caption{Average area of a useful monolayer at each parameter.}
    \begin{tabular}{|c|c|c|c|c|c|}
    \hline
        $\langle A_{UML}\rangle$ &  1 $\mu$m/s& 10 $\mu$m/s &  100 $\mu$m/s&1 cm/s& Mean \\
        ($\mu$m$^2$/flake)&&&&&\\
        \hline
        25$^{\circ}$C & 430$\pm$180 & 370$\pm$50 & 490$\pm$210 & 470$\pm$210 & 430$\pm$160\\
         \hline
         65$^{\circ}$C & 470$\pm$180 & 410 & 370$\pm$50 & 440&440$\pm$130\\
         \hline
         105$^{\circ}$C & 460 & 380 & 440&360&420$\pm$50\\
         \hline
         Mean&460$\pm$150&380$\pm$50&460$\pm$180&440$\pm$170&430$\pm$140\\
         \hline
    \end{tabular}
    \label{tab:resultstable6}
\end{table}
As shown in the preceding tables, the standard deviations of most measured metrics are comparable to the mean observed values; though this precludes strong conclusions about the dependence of yield on peel speed or anneal temperature, some weak trends may still be observed. This is clarified by calculating Pearson's product-moment correlation coefficient $r$, which describes the strength of linear associations between two variables - true positive (negative) linear dependence is represented by $r=1(-1)$, with a total lack of dependence represented by $r=0$. The absolute values of $|r|\approx0.25$ indicate a weak relationship between peel speed and normalized UML yield, with only 6\% of the variation in yield explained by variation in peel speed. We suspect the remaining variation mainly derives from inherent variance in the properties of tapes manually populated with natural graphite, though other uncontrolled factors such as ambient humidity or substrate contamination may also contribute.
\begin{table}[h]
    \centering
    \caption{Pearson product-moment correlation coefficients $r$ between controlled variables $T,V$ and the respective dependent variables.}
    \begin{tabular}{|c|c|c|c|c|c|}
    \hline
        $r$ &  $\langle N_{UML} \rangle$& UML/A$_{gr}$ &  A$_{UML}$/A$_{gr}$& UML Yield&$\langle A_{UML}\rangle$\\
        &(N/wafer)&(N/mm$^2$)&($\mu$m$^2$/mm$^2$)&(\%)&($\mu$m$^2$/flake)\\
        \hline
        T & 0.18 & 0.13 & 0.07 & -0.08&-0.02\\
         \hline
         V& -0.34 & -0.32 & -0.28&-0.24&0.001\\
         \hline
    \end{tabular}
    \label{tab:suppcorrelations}
\end{table}
Though at slower peel speeds and 65$^\circ$C monolayer yield per wafer seems to somewhat increase, the concomitant increase in residue and added (unattended) time per exfoliation are unfavorable. Thus, going forward we would prefer higher peel speeds and lower temperature anneals. Aside from the annealing and peeling steps, typical sample preparation for the eXfoliator takes approximately 15 minutes, with another 60 minutes post-exfoliation to scan and search each wafer.
For comparison, we surveyed our group members with varying levels of experience with manual exfoliation, asking them how long they took to exfoliate graphene over their preferred substrate area, and how long they took to then identify all useful monolayers. By naive scaling of the survey results, we conclude that a typical user would take between 25 and 150 minutes to manually exfoliate the area produced by a single run of the eXfoliator, followed by a manual searching time of 30 to 120 minutes.

\begin{figure}[h]
    \centering
    \includegraphics[width=0.42\textwidth]{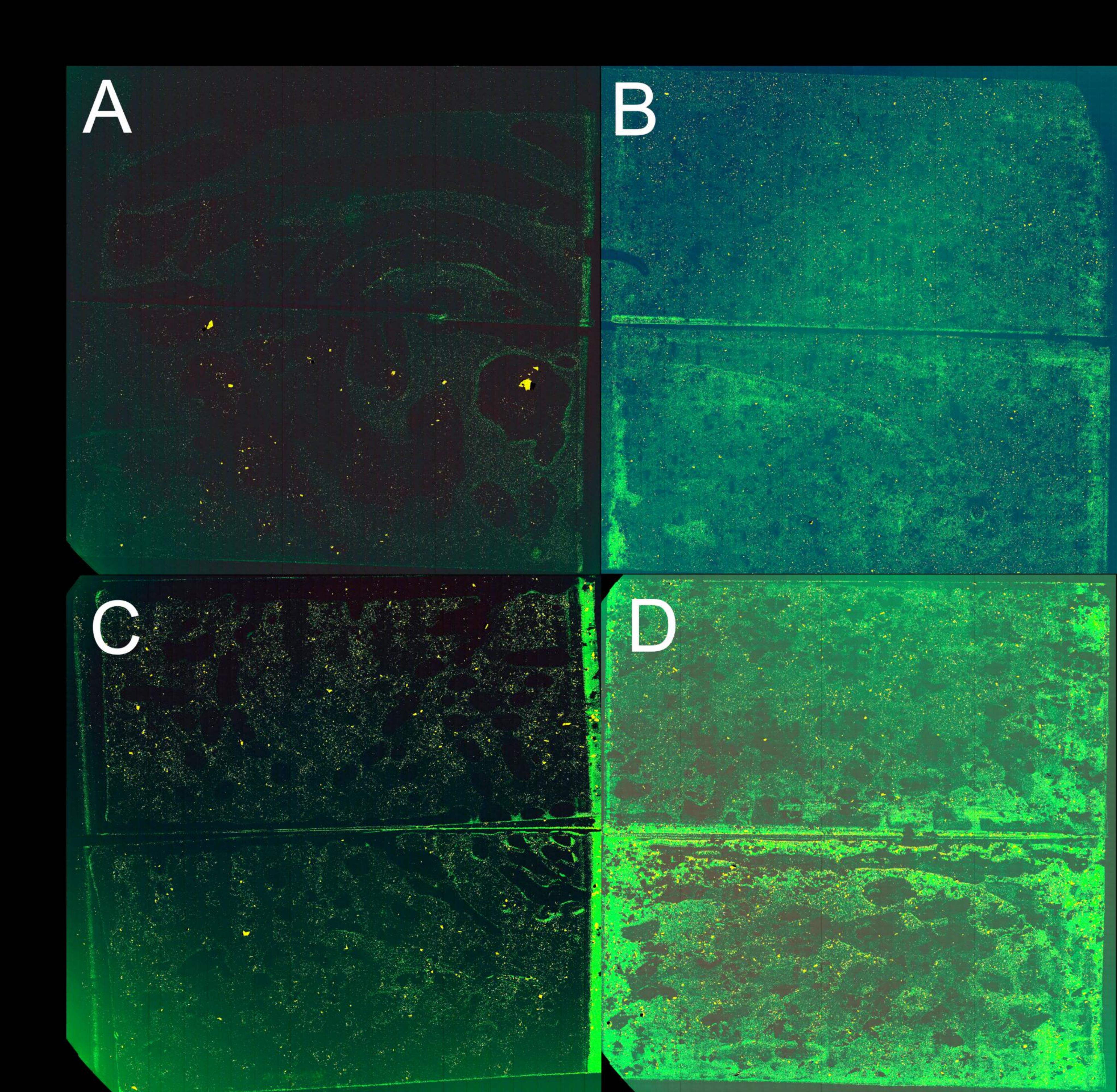}
    \caption{Wafer-scale images of samples post-exfoliation, in false color to highlight tape residue (green), with thicker residue showing as a more intense color. Anneal temperatures and peel speeds were A. 25$^{\circ}$C, 1 cm/s, B. 105$^{\circ}$C, 1 cm/s, C. 25$^{\circ}$C, 100 $\mu$m/s, and D. 105$^{\circ}$C, 100 $\mu$m/s. Each image was taken under identical lighting settings.}
    \label{fig:residue}
\end{figure}

In practice, as the eXfoliator can be prepared and run in parallel with the mapping microscope, maximum sample production is in practice reached by all recipes that take less than the 45 minutes the microscope needs to map the relevant area.
With the approximate time to complete each recipe given by Table \ref{tab:suppspeedtable}, this limits efficient exfoliation recipes to peel speeds of 100 $\mu$m/s or higher. As there is no strong correlation between individual monolayer area and anneal temperature or peel speed, the figure of merit for this study is taken to be number of monolayers per starting area of graphite; as shown in Table \ref{tab:resultstable3}, annealing at 65$^{\circ}$C seems to yield better results than the other temperatures.
However, because tape adhesive is a viscous polymer that readily flows at these temperatures, slower peel speeds and higher anneal temperatures result in increased levels of polymer residue on the wafer, as shown in Figure \ref{fig:residue}.
Monolayers are primarily cleaved from the bottom of larger flakes and are thus often pristine regardless of the amount of residue elsewhere on the wafer, but stamp-based stacking techniques are often hindered by the presence of thick tape residue nearby.
We have no solvent-free technique for reliably removing heavy polymer residue left by Scotch magic tape without damaging flakes, so it is desirable to use exfoliation recipes that minimize tape residue while retaining sufficient yield of flakes.
With the constraints of time and residue, among the recipes explored in this study we identified annealing at 25$^\circ$C and peeling at 100 $\mu$m/s to be the most promising for future use.

Overall, approximately 34\% of all monolayers detected with areas greater than 200 $\mu$m$^2$ were freestanding, useful monolayers. As shown in Appendix C, the total and useful monolayer yield per wafer of most parameter pairs did not significantly deviate from the average, suggesting peel speed and anneal temperature are not the dominant causes of flake area distribution. The effect of uncontrolled variables, such as humidity and initial graphite thickness, was not examined in this study, but the eXfoliator design allows for further investigation if desired.
\begin{figure}[h]
    \centering
    \includegraphics[width=0.45\textwidth]{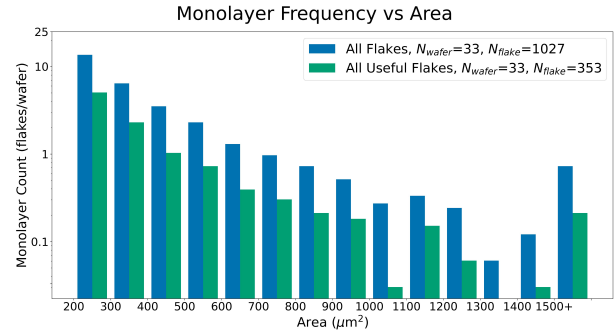}
    \caption{Monolayer frequency vs area, with 100 $\mu$m$^2$ bins. The distribution of useful monolayers (green) is shown against that of all monolayers detected (blue), with a final proportion of 353/1027. Parameter-wise histograms are given in Appendix C.}
    \label{fig:freevstotal}
\end{figure}

\section{Conclusions}
This paper introduces a simple, compact device for semi-automated exfoliation of monolayer materials, and describes the imaging system used to characterize the performance of the device across a range of parameters.
Overall, the eXfoliator system successfully produces sufficient numbers of large, freestanding monolayers for use in device fabrication, with an output of approximately three large monolayers per wafer. The eXfoliator appears to improve reproducibility and throughput compared to manual exfoliation, but to robustly validate this assessment further studies of flake yield with significantly more samples or a more reproducible technique for populating the initial tape would be needed. The advantage of the eXfoliator may be primarily in allowing beginning users to achieve a rate of flake production they would otherwise take months to reach.
Unlike Brookhaven's QPress, the eXfoliator isn't presently operated in an inert gas atmosphere, but its small footprint allows for placement into a glove-box if exfoliation of air-sensitive materials is desired. Compared to simpler automated peeling setups, the eXfoliator allows for control of peel angle and application pressure\citep{Wang2022email}. As the application pressure may affect substrate-crystal adhesion, especially in the presence of atmospheric adsorbates such as water and hydrocarbons, and the peel angle is likely to affect the ratio of shear to normal forces in the crystal during peeling, these parameters are deemed relevant to control.

We find that flake yield from Scotch tape exfoliation does not strongly depend on anneal temperature and peel speed over the range of parameter values investigated. In practice, the production rate of useful monolayers was most heavily affected by the total time to exfoliate each sample wafer and the amount of polymer tape residue remaining on an exfoliated wafer.

The eXfoliator's design is well-suited for production of other van der Waals materials, such as hexagonal boron nitride and transition metal dichalcogenides, and allows for further exfoliation studies using other substrates, materials, and parameters. Limitations of the existing eXfoliator design which could be addressed include its slow cooling rate, placing a lower limit on the tape-silicon dwell time if heating is used, and lack of rotational control in the Y-Z plane, so populated tape cannot be ''rolled on'' to silicon substrates.

\section{Acknowledgments}
The authors thank Chaitrali Duse and Steven Tran for their aid in preparing several samples; Joe Finney, Rupini Kamat, and Greg Zaborski Jr. for their insight and feedback in designing a machine to replicate manual exfoliation of monolayer graphene; Andy Mannix for useful discussions about the exfoliation process; and Feng Wang, Jingxu Xie, and Emma Regan for useful discussions about the reasoning behind their exfoliator's design. Experimental measurements and analysis, parts for building the automated exfoliator, and acquisition of the motorized microscope for efficient imaging of the output of the exfoliator, were supported by the U.S. Department of Energy, Office of Basic Energy Sciences (DE-SC0021984).
Supporting infrastructure was funded in part by the Gordon and Betty Moore Foundation through Grant No. GBMF3429 and Grant No. GBMF9460.
Sandia National Laboratories is a multimission laboratory managed and operated by National Technology and Engineering Solutions of Sandia, LLC., a wholly owned subsidiary of Honeywell International, Inc., for the U.S. Department of Energy’s National Nuclear Security Administration under contract DE-NA-0003525.

The authors have no conflicts to disclose.
The data that support the findings of this study are available from the corresponding author upon reasonable request.
The authors contributed to this manuscript in the following ways -
Elijah Courtney: Conceptualization (equal); Data Curation; Formal Analysis (lead); Investigation; Methodology (lead); Software (support); Original Draft Preparation; Review and Editing (equal). Mihir Pendharkar: Conceptualization (equal); Methodology (support); Review and Editing (equal). Nathan Bittner: Conceptualization (support); Software (lead); Review and Editing (equal). Aaron Sharpe: Conceptualization (support); Formal Analysis (support); Methodology (support); Review and Editing (equal). David Goldhaber-Gordon: Conceptualization (support); Methodology (support); Project Administration; Supervision; Review and Editing (equal).
\section{Appendix A - eXfoliator Parts}
\begin{enumerate}[noitemsep]
  \item Aluminum 1515 t-slot frame - 4$\times$ 12" rails, 10$\times$ 18" rails, with appropriate right-angle brackets
  \item Aluminum VT-50L mount, custom machined
  \item Translation Stage VT-50L, 200mm travel, via MicronixUsa
  \item Aluminum XZ bracket, custom machined
  \item Aluminum crossbeam, custom machined
  \item Flanged ball bearing carriage, via McMaster-Carr (6709K11)
  \item 340mm carriage guide rail, via McMaster-Carr (6709K33)
  \item Guide rail mount for t-slot frame, via McMaster-Carr (1748N14)
  \item Translation Stage VT-50L, 100mm travel, via MicronixUsa
  \item Aluminum standoff, custom machined
  \item Aluminum upper post, custom machined
  \item Aluminum load cell mount, custom machined
  \item LCF300 25lb universal load cell, via FUTEK (FSH04274) with USB cable (FSH04720, FSH04742)
  \item Aluminum lower post, custom machined
  \item Aluminum post mount, custom machined
  \item Aluminum tape plate, custom machined
  \item RET Control Visc hot plate, via IKA (0005020001)
  \item Aluminum plate adapter, custom machined
  \item Aluminum wafer pedestal, custom machined
  \item Aluminum wafer pedestal top, custom machined
  \item[Note:] The hot plate's firmware was insensitive to changes in weight below 2g. The design relied on making gradual contact with live feedback from the weight sensor, so the hot plate could not be used for the intended purpose of providing both digital temperature control and force feedback, inspiring the load cell's incorporation.
  \item Aluminum temperature probe sabot, custom machined
  \item Aluminum mounting plate, via McMaster-Carr
\end{enumerate}
\vfill\eject

\section{Appendix B - Additional Tables}
\label{sec:tabappendix}
\begin{table}[h]
    \centering
        \caption{Number of sample wafers exfoliated at each peel speed and anneal temperature, with application pressure, dwell time, peel angle, and peel temperature held constant.}
    \begin{tabular}{|c|c|c|c|c|c|}
    \hline
         Count (N)&  1 $\mu$m/s& 10 $\mu$m/s &  100 $\mu$m/s&1 cm/s& Total \\
        \hline
        25$^{\circ}$C & 2 &7 & 8 & 4&21\\
         \hline
         65$^{\circ}$C & 4 & 1 & 2 & 1&8\\
         \hline
         105$^{\circ}$C & 1 & 1 & 1&1&4\\
         \hline
         Total&7&9&11&6&33\\
         \hline
    \end{tabular}
    \label{tab:suppresultstable1}
        \caption{Number of useful monolayers per sample at each peel speed and anneal temperature, presented with standard deviations.}
    \begin{tabular}{|c|c|c|c|c|c|}
    \hline
         $\langle N_{UML} \rangle$&  1 $\mu$m/s& 10 $\mu$m/s &  100 $\mu$m/s&1 cm/s& Mean \\
         (N/wafer)&&&&&\\
        \hline
        25$^{\circ}$C & 10$\pm$2 & 13$\pm$3& 9$\pm$5 & 5$\pm$3&10$\pm$5\\
         \hline
         65$^{\circ}$C & 14$\pm$4 & 17 & 9.5$\pm$ 0.7 & 16&14$\pm$4\\
         \hline
         105$^{\circ}$C & 10 & 8 & 16&8&11$\pm$4\\
         \hline
         Mean&12$\pm$4&13$\pm$4&10$\pm$5&8$\pm$5&11$\pm$5\\
         \hline
    \end{tabular}
    \label{tab:suppresultstable2}
        \caption{Number of useful monolayers per square mm of graphite on the initial tape at each peel speed and anneal temperature, presented with standard deviations.}
    \begin{tabular}{|c|c|c|c|c|c|}
    \hline
         UML/A$_{gr}$ &  1 $\mu$m/s& 10 $\mu$m/s &  100 $\mu$m/s&1 cm/s& Mean \\
         (N/mm$^2$)&&&&&\\
        \hline
        25$^{\circ}$C & 1.7$\pm$0.1 &2.4$\pm$1.6 & 1.4$\pm$0.9 & 0.7$\pm$1.2&1.6$\pm$1.3\\
         \hline
         65$^{\circ}$C & 3.4$\pm$0.8 & 2.9 & 2.6$\pm$1 & 1.8&2.8$\pm$0.9\\
         \hline
         105$^{\circ}$C & 2.9 & 0.9 & 1.6&2.2&1.6$\pm$0.9\\
         \hline
         Mean&2.8$\pm$1&2.2$\pm$1.5&1.5$\pm$1&1.1$\pm$1&1.8$\pm$1.3\\
         \hline
    \end{tabular}
    \label{tab:suppresultstable3}
\end{table}
\FloatBarrier
The normalized counts reported in Table \ref{tab:suppresultstable3} were multiplied by the average area of useful monolayers produced at the relevant parameters. As shown in Table \ref{tab:suppresultstable4}, slower peel speeds tend to yield a larger total area of monolayer graphene, but large standard deviations prevent any strong conclusions from being drawn.
\begin{table}[h]
    \centering
        \caption{Total area of useful monolayers per square mm of graphite on the initial tape at each peel speed and anneal temperature, presented with standard deviations.}
    \begin{tabular}{|c|c|c|c|c|c|}
    \hline
         A$_{UML}$/A$_{gr}$ &  1 $\mu$m/s& 10 $\mu$m/s &  100 $\mu$m/s&1 cm/s& Mean \\
         ($\mu$m$^2$/mm$^2$)&&&&&\\
        \hline
        25$^{\circ}$C & 740$\pm$240 &900$\pm$730 & 680$\pm$870 & 340$\pm$520&680$\pm$730\\
         \hline
         65$^{\circ}$C & 1620$\pm$810 & 1190 & 970$\pm$220 & 800&1250$\pm$670\\
         \hline
         105$^{\circ}$C & 1330 & 350 & 690&800&680$\pm$410\\
         \hline
         Mean&1280$\pm$740&840$\pm$690&710$\pm$740&480$\pm$440&790$\pm$700\\
         \hline
    \end{tabular}
    \label{tab:suppresultstable4}
\end{table}

Table \ref{tab:suppresultstable5} shows the percentage of all detected monolayers (including monolayers smaller than 200 $\mu$m$^2$ and those attached to bulk) that were useful monolayers, with slower peel speeds and an anneal temperature of 65$^\circ$C tending to yield a higher proportion of useful monolayers.
\begin{table}[h]
    \centering
    \caption{Yield of useful monolayers as a percentage of all detected monolayers, at each peel speed and anneal temperature, presented with standard deviations.}
    \begin{tabular}{|c|c|c|c|c|c|}
    \hline
        UML Yield (\%) &  1 $\mu$m/s& 10 $\mu$m/s &  100 $\mu$m/s&1 cm/s& Mean \\
        \hline
        25$^{\circ}$C & 15$\pm$7& 10$\pm$4 & 9$\pm$3 & 10$\pm$7 &10$\pm$5\\
         \hline
         65$^{\circ}$C & 19$\pm$5 & 19.8 & 13$\pm$0.4 & 9.5&15$\pm$5\\
         \hline
         105$^{\circ}$C & 19.2 & 8.6 & 6.3&14&9$\pm$6\\
         \hline
         Mean&18$\pm$5&10$\pm$5&9$\pm$3&10$\pm$6&11$\pm$5\\
         \hline
    \end{tabular}
    \label{tab:suppresultstable5}
\end{table}
\begin{table}[h]
    \centering
        \caption{Average area of a useful monolayer at each peel speed and anneal temperature.}
    \begin{tabular}{|c|c|c|c|c|c|}
    \hline
        $\langle A_{UML}\rangle$ &  1 $\mu$m/s& 10 $\mu$m/s &  100 $\mu$m/s&1 cm/s& Mean \\
        ($\mu$m$^2$/flake)&&&&&\\
        \hline
        25$^{\circ}$C & 430$\pm$180 & 370$\pm$50 & 490$\pm$210 & 470$\pm$210 & 430$\pm$160\\
         \hline
         65$^{\circ}$C & 470$\pm$180 & 410 & 370$\pm$50 & 440&440$\pm$130\\
         \hline
         105$^{\circ}$C & 460 & 380 & 440&360&420$\pm$50\\
         \hline
         Mean&460$\pm$150&380$\pm$50&460$\pm$180&440$\pm$170&430$\pm$140\\
         \hline
    \end{tabular}
    \label{tab:suppresultstable6}
        \caption{Approximate time to anneal and peel a sample at each peel speed and anneal temperature, in minutes.}
    \begin{tabular}{|c|c|c|c|c|c|}
    \hline
        Time &  1 $\mu$m/s& 10 $\mu$m/s &  100 $\mu$m/s&1 cm/s\\
        (min/wafer)&&&&\\
        \hline
        25$^{\circ}$C & 800 & 80 & 8 & 0.1\\
         \hline
         65$^{\circ}$C & 815 & 95 & 23&15\\
         \hline
         105$^{\circ}$C & 825 & 105 & 33&25\\
         \hline
    \end{tabular}
    \label{tab:suppspeedtable}
\end{table} \FloatBarrier
\section{Appendix C - Additional Figures}
\begin{figure}[h]
    \centering
    \includegraphics[width=0.45\textwidth]{Images/Overall_plot.pdf}
    \caption{A histogram of flake area, identical to that shown in Fig.\ref{fig:freevstotal}. The distribution of useful monolayers (blue) is shown against that of all monolayers detected (green), with 100 $\mu$m$^2$ bins and a final proportion of 353/1027.}
    \label{fig:freevstotal_2}
    \includegraphics[width=0.45\textwidth]{Images/templot_total.pdf}
    \label{fig:Tempplot_a}
    \caption{The distributions of monolayer frequency (flakes/wafer) on wafers annealed at 25$^\circ$C (green), 65$^\circ$C (orange), and 105$^\circ$C (pink) are shown against that of all monolayers found (blue), with 100 $\mu$m$^2$ bins.}
    \includegraphics[width=0.45\textwidth]{Images/tempplot_useful.pdf}
    \label{fig:Tempplot_u}
    \caption{The distributions of useful monolayer frequency (flakes/wafer) on wafers annealed at 25$^\circ$C (green), 65$^\circ$C (orange), and 105$^\circ$C (pink) are shown against that of all useful monolayers found (blue), with 100 $\mu$m$^2$ bins.}
\end{figure}
\begin{figure}[h]
    \centering
    \includegraphics[width=0.45\textwidth]{Images/speedplot_total.pdf}
    \label{fig:speedplot_a}
    \caption{The distributions of monolayer frequency (flakes/wafer) on wafers peeled at 1 $\mu$m/s (green), 10 $\mu$m/s (orange), 100 $\mu$m/s (pink), and 1 cm/s (yellow) are shown against that of all monolayers found (blue).}
    \includegraphics[width=0.45\textwidth]{Images/speedplot_useful.pdf}
    \label{fig:speedplot_u}
    \caption{The distributions of useful monolayer frequency (flakes/wafer) on wafers peeled at 1 $\mu$m/s (green), 10 $\mu$m/s (orange), 100 $\mu$m/s (pink), and 1 cm/s (yellow) are shown against that of all useful monolayers found (blue).}
\end{figure}
\begin{figure}[h]
    \centering
    \includegraphics[width=0.45\textwidth]{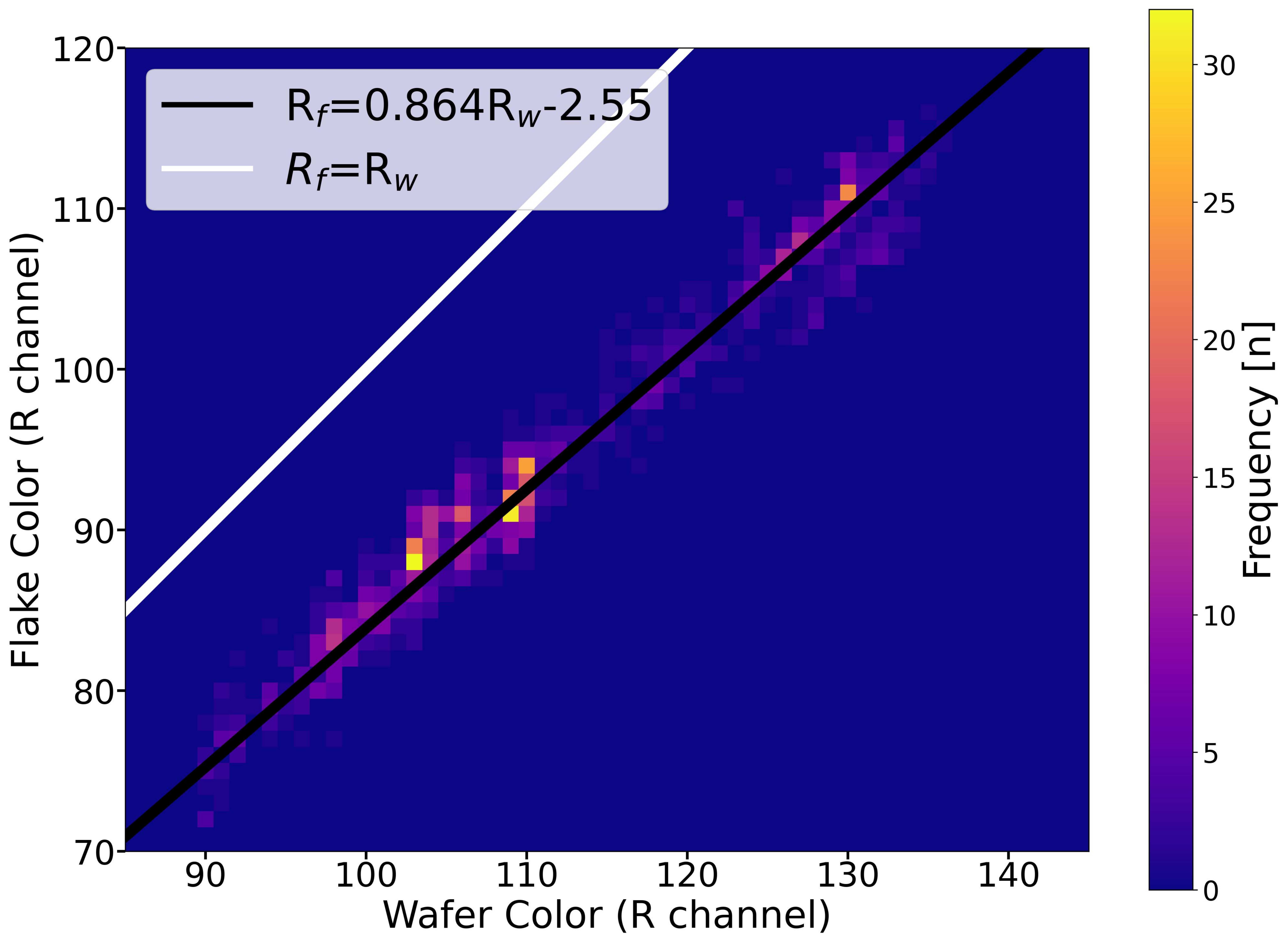}
    \label{fig:rfvsrw}
    \includegraphics[width=0.45\textwidth]{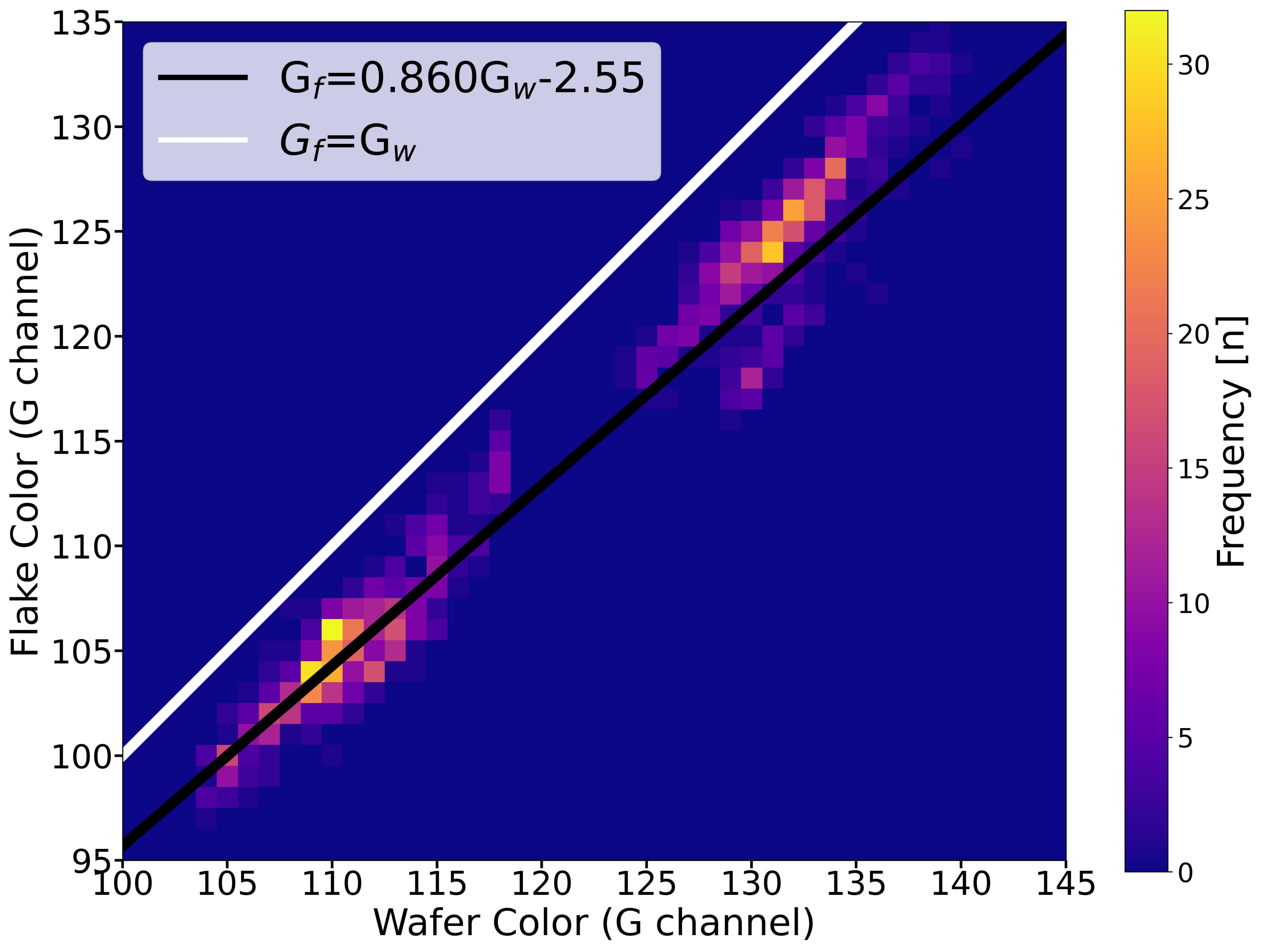}
    \label{fig:gfvsgw}
    \includegraphics[width=0.45\textwidth]{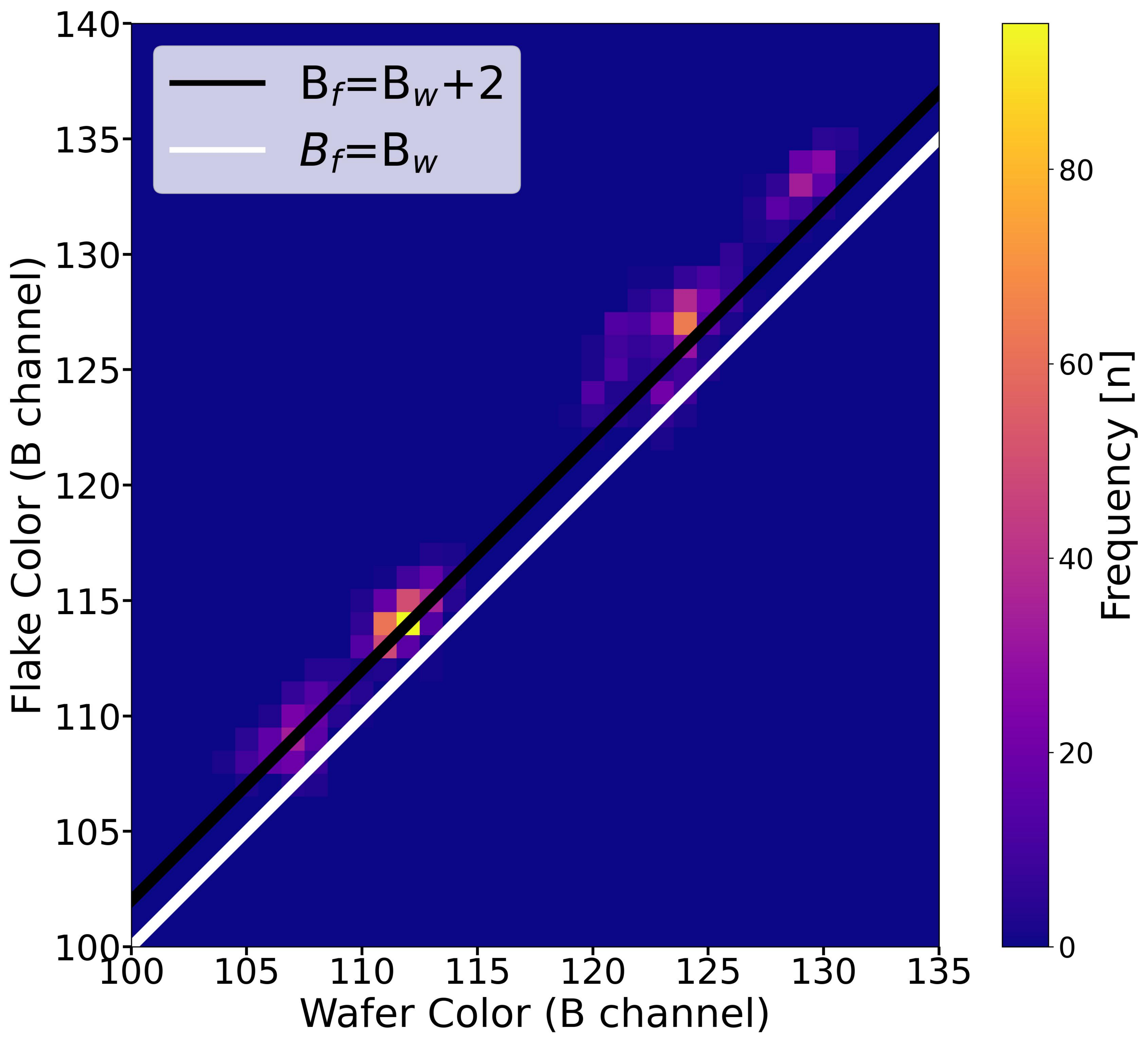}
    \caption{Frequency of detected flake color vs local background color in the red, green, and blue channels. The predictive empirical functions R$_f$=0.864R$_w$-2.55, G$_f$=0.860G$_w$+9.68, and B$_f$=B$_w$+2 were used, with all pixels within Euclidean distance 8 of the predicted flake RGB passed to the cluster-finding portion of the algorithm. The apparent bimodality is due to batch differences between the two wafer cassettes used in this experiment.}
    \label{fig:bfvsbw}
\end{figure}
\FloatBarrier
\bibliography{references}

\end{document}